\documentclass[submission]{eptcs}
\providecommand{\event}{ICE 2012}

\def\finex{{\unskip\nobreak\hfil
\penalty50\hskip1em\null\nobreak\hfil$\diamond$
\parfillskip=0pt\finalhyphendemerits=0\endgraf}}

\newcounter{mmyexample} 
\newenvironment{example}{%
  \refstepcounter{mmyexample}%
  {\par\bigskip\noindent\textbf{Example~\arabic{mmyexample}}.%
  }%
}{\finex\bigskip}

\newcounter{myproposition} 
\newenvironment{proposition}{%
  \refstepcounter{myproposition}%
                 {\par\bigskip\noindent\textbf{Proposition~\arabic{myproposition}}.
                 }
}{\bigskip}

\newcounter{mytheorem} 
\newenvironment{theorem}{%
  \refstepcounter{mytheorem}%
  {\par\bigskip\noindent\textbf{Theorem~\arabic{mytheorem}}.}
}              

\usepackage{graphicx}
\usepackage{amsmath}
\usepackage{amsthm}
\usepackage{amssymb}
\usepackage{pdfsync}
\usepackage[usenames,dvipsnames]{pstricks}
\usepackage{epsfig}
\usepackage{pst-grad} 
\usepackage{pst-plot} 
\usepackage[arrow,matrix,frame,curve]{xy}
\usepackage{color,graphics,xcolor}
\usepackage{ulem}
\usepackage{hyperref}

\title{Enforcing Architectural Styles in Presence of \\ Unexpected
  Distributed Reconfigurations\thanks{This work has been supported by FP7-PEOPLE-2011-IRSES MEALS}}

\author{Kyriakos Poyias \qquad \qquad Emilio Tuosto
  \institute{Department of Computer Science, University of Leicester,
    UK} \email{kyriakos@le.ac.uk \qquad \qquad \quad emilio@le.ac.uk}
}

\def\titlerunning{Enforcing Architectural Styles in Presence of
  Unexpected Reconfigurations} \def\authorrunning{K. Poyias,
  E. Tuosto}

\newcommand{\hidden}[1]{}
\newcommand{\lvar}[1]{{\mathsf{#1}}}

\newcommand{\wpre}[4]{\mathcal{W}^{#1}_{#2}(#3,#4)}

  \definecolor{shadecolor}{rgb}{1,0.99,0.9}

  \definecolor{greenshade}{RGB}{193,255,193}

\newcommand{\curlybraces}[1]{\lbrace #1 \rbrace}

\newcommand{\anglebraces}[1]{\langle #1 \rangle}

\newcommand{\eqname}{\mathit{eq}}
\renewcommand{\wp}[3]{\mathit{wp}^{#1}_{#2}({#3})}
\newcommand{\fwp}[4]{\eqname^{{#1},{#2}}_{{#3}}({#4})}
\newcommand{\wDef}[3]{\mathit{wd}^{#1}_{#2}({#3})}
\newcommand{\wpf}[2]{\mathit{wp}^{{#1},{#2}}}

\newcommand{\wDeff}[2]{\mathit{wd}^{#1}_{#2}}
\renewcommand{\choose}[3]{\tilde{#1} \ \text{on}\ {#2}\cdot{#3}}
\newcommand{\envirWp}{\mathcal{E}}
\newcommand{\envir}[2]{\mathcal{E}^{(#2)}(#1)}
\newcommand{\envirNot}[4]{\mathcal{E}({#1}) \stackrel{as}{=} {#4}\;{#2}\;{#3}}
\newcommand{\envempty}{\mathbf{0}}

\newtheorem{mydef}{Definition}

\newcommand{\nodes}[1]{V_{#1}}
\newcommand{\edges}[1]{E_{#1}}
\newcommand{\tentacle}[1]{t_{#1}}
\newcommand{\fv}[1]{\mathrm{fv}{(#1)}}

\newcommand{\true}{\top}
\newcommand{\false}{\bot}

\newcommand{\tg}{\Gamma}
\newcommand{\varset}{\mathsf{V}}
\newcommand{\nodeset}{\mathfrak{N}}
\newcommand{\edgeset}{\mathfrak{E}}
\renewcommand{\L}{\mathcal{L}}
\newcommand{\typing}[1]{\tau_{#1}}
\newcommand{\subs}[2]{[#1 \mapsto #2 ]}
\newcommand{\tsubs}[2]{\subs{\tilde{#1}}{\tilde{#2}}}
\newcommand{\mmdef}{\mbox{$\;\stackrel{\mathrm{def}}{=}\;$}}
\newcommand{\inodes}[1]{{#1}^\circ}
\newcommand{\st}{\big|}

\newcommand{\set}[1]{{\curlybraces{{#1}}}}
\newcommand{\tentfun}[3]{{#1} : {#2} \mapsto ({#3})}

\newcommand{\conf}[1]{\langle {#1} \rangle}
\newcommand{\no}[1]{\mathtt{noEdge}\conf{#1}}
\newcommand{\pa}[2]{\mathtt{path}\ {#1} \ {#2}}
\newcommand{\after}[2]{{#1} \circ {#2}}

\newcommand{\prd}[3]{\set{#1}\ {#2} \ \set{#3}}

\newcommand{\accessport}{\circ}
\newcommand{\chainport}{\bullet}
\newcommand{\tedge}[1]{*+[F]{{#1}}}
\newcommand{\ntedge}[1]{*+[F=]{{#1}}}

\newcommand{\classes}{C}

\newcommand{\ntaritythreeT}[4]{\ntedge{{#1}} \ar[{#2}] \ar@{-}[{#3}] \ar@{-}[{#4}]}
\newcommand{\ntaritythree}[5]{\ntedge{{#1}} \ar@(r,l)[{#2}] \ar@{-}@(d,{#3})[{#4}] \ar@{-}@(l,r)[{#5}]}
\newcommand{\ntaritytwo}[3]{\ntedge{{#1}} \ar[{#2}] \ar@{-}[{#3}]}
\newcommand{\ntarityone}[2]{\ntedge{{#1}} \ar[{#2}]}
\newcommand{\taritythree}[4]{\tedge{{#1}} \ar@(r,l)[{#2}] \ar@{-}@(d,u)[{#3}] \ar@{-}@(l,r)[{#4}]}
\newcommand{\taritytwo}[3]{\tedge{{#1}} \ar@(r,l)[{#2}] \ar@{-}@(l,r)[{#3}]}
\newcommand{\tarityone}[2]{\tedge{{#1}} \ar[{#2}]}

\newcommand{\client}{\texttt{C}}
\newcommand{\clientobj}[1]{{{#1}:\client}}

\newcommand{\bookF}{\texttt{BF}}

\newcommand{\findF}{\texttt{FF}}
\newcommand{\findFobj}[1]{{{#1}:\findF}}
\newcommand{\flights}{\texttt{Fls}}
\newcommand{\flightsobj}[1]{{{#1}:\flights}}
\newcommand{\flight}{\texttt{Fl}}
\newcommand{\flightobj}[1]{{{#1}:\flight}}
\newcommand{\pay}{\texttt{P}}
\newcommand{\payobj}[1]{{{#1}:\pay}}
\newcommand{\fpay}{\texttt{PF}}
\newcommand{\fpayobj}[1]{{{#1}:\fpay}}

\newcommand{\findflightprod}{\mathtt{findFlights}}
\newcommand{\bookflightprod}{\mathtt{bookFlight}}
\newcommand{\browseflightsprod}{\mathtt{browseFlights}}
\newcommand{\noFsprod}{\mathtt{noFlights}}
\newcommand{\delFprod}{\mathtt{deleteFlight}}

\newcommand{\chainprod}{\bookflightprod}

\begin{document}

\maketitle

\begin{abstract}
  Architectural Design Rewriting (ADR, for short) is a rule-based
  formal framework for modelling the evolution of architectures of
  distributed systems.
  Rules allow ADR graphs to be refined.
  After equipping ADR with a simple logic,
  we equip rules with pre- and post-conditions;
  the former constraints the applicability of the rules
  while the later specifies properties of the resulting graphs.
  We give an algorithm to compute the weakest pre-condition out of
  a rule and its post-condition.
  On top of this algorithm, we design a simple methodology that allows
  us to select which rules can be applied at the architectural level
  to reconfigure a system so to regain its architectural style when it
  becomes compromised by unexpected run-time reconfigurations.

\end{abstract}

\section{Introduction}
\label{sec:intro}
Modern applications are very rarely developed as ``stand-alone''
software; as a matter of fact, even simple applications are nowadays
\emph{open} in the sense that they are typically able to connect
and/or be integrated with other applications such as those in
service-oriented or cloud computing.
Also, this kind of software tend to be \emph{autonomic}, namely it
needs to automatically adapt to the (often unpredictable)
run-time changes.

Openness magnifies the complexity of such software.
In fact, open systems are subject to unexpected reconfigurations that
may hinder their execution and drive computations into erroneous
states in an unanticipated manner.
Detecting and tackling those states of the computation at run-time is
crucial to re-establish correct configurations from which the
computation can safely restart.
For example, the reaction to the failure of a service $S$, may
redirect the requests of the clients to another service $S'$.

A problem that can arise in those cases is that the run-time
reconfigurations may compromise the alignment with the expected
abstract architecture.
In the client-service scenario mentioned above, the choice of $S'$ may
cause the violation of some architectural constraints designed e.g. to
balance the load.

In this paper we propose to use high-level designs of software
architectures to drive system reconfigurations so that desirable
architectural properties (expressed as logical invariants) are
maintained when reconfigurations are necessary.
Software architectures specify the structure and interconnections
of a software product.
Ordinary computation can change the state, but they are very rarely
allowed to modify the architecture.
In this context it is also crucial to preserve \emph{architectural
  styles}~\cite{Garlan:Perspectives} that allow one ($i$) to specify
(reusable) design patterns, ($ii$) to confine the parts to be
reconfigured, and ($iii$) to control the architectural changes.

Our approach hinges on a formal language for specifying software
architectures, their refinements, and their style.
Methodologically, we adopt ADR~\cite{bllmt08} as our architectural
description language.
As surveyed in \S~\ref{sec:adr}, ADR models systems as
\emph{(hyper)graphs} that is a set of \emph{(hyper)edges} sharing some
nodes; respectively, edges represent distributed components (at some
level of abstraction) while nodes represent communication ports.
Also, ADR features refinement rules of the form $L \to R$ where 
$L$ is a (hyper)edge and $R$ a (hyper)graph meant to replace $L$ with
$R$ within a given graph.
In ADR, a system corresponds to a configuration of elements
(i.e. nodes and edges)
that can be related to the architecture graph
components and expected to respect the architectural style specified by
the refinement rules.
Such elements can interact through their connections according to
run-time interactions (run-time reconfigurations) not
represented at the architectural level.
A main reason for adopting ADR is that it has been designed to
support the alignment of architecture-related information with
run-time behaviour in order to drive execution.

A technical contribution of this paper (\S~\ref{sec:logic} and
\S~\ref{sec:dbc}) is to generalise ADR with
\emph{asserted productions}, that is refinement rules of the form
\begin{equation}
  \set \psi L \to R \set \varphi \qquad \text{where}\ \psi
  \ \text{and}\ \varphi \ \text{are the pre- and post-conditions,
    respectively}
\label{eq:asserted}
\end{equation}
The intuition is that~\eqref{eq:asserted} can be applied only to
graphs satisfying $\psi$ to obtain a graph satisfying $\varphi$.
For this, we use a simple logic for hyper graphs.

In ADR, architectural styles are formalised in terms of productions that
describe the legal configurations of systems.
We generalise this by envisaging architectural styles as set of
productions together with invariants (expressed as closed formulae of
our logic) which can be thought of as \emph{contracts} that
architectures have to abide by.

The main result of the paper is an algorithm (\S~\ref{sec:wpc}) to
compute the \emph{weakest} pre-condition from the post-condition of a
production.
Also, we use such algorithm to devise a methodology to re-establish
the architectural style specified for a system when run-time
reconfigurations compromise it.

\newcommand{\secref}[1]{~\S~\ref{#1}}

\paragraph{Synopsis}
A short overview of ADR is given in\secref{sec:adr} (for simplicity,
we do not describe ADR reconfiguration; the interested reader is
referred e.g. to~\cite{bllmt08} for the technical details).
We introduce a simple logic for ADR in\secref{sec:logic}.
Basic definitions to specify our algorithm are in\secref{sec:dbc}
while the algorithm is in\secref{sec:wpc}.
In\secref{sec:methodology} we describe a methodology that relies on
the algorithm in\secref{sec:wpc} to recover architectural styles
compromised by run-time reconfigurations.
An application of the methodology is given in\secref{sec:casestudy}.
Related work are discussed in\secref{sec:rw}.
Concluding remarks and future work is in \secref{sec:conc}.


\section{A walk through ADR}
\label{sec:adr}
We briefly overview ADR; we borrow from~\cite{bllmt08} the main
definitions and notations (slightly adapting them to our needs).

In the following, $\nodeset$ and $\edgeset$ are two countably infinite
and disjoint sets (of nodes and edges respectively), $X^*\mmdef
\set{(x_1,\ldots,x_n) \st x_1,\ldots,x_n \in X}$ is the set of finite
lists on a set $X$, and $\tilde x$ ranges over $X^*$.
Also, abusing notation, we sometimes use $\tilde x$ to
indicate its underlying set of elements.
\begin{mydef}[(Hyper)graphs]\label{graphsDef}
  A \emph{(hyper)graph} is a tuple $G = \anglebraces{V, E, t}$
  where $V \subseteq \nodeset$ and $E \subseteq \edgeset$ are finite
  and $t : E \to V^*$ is the \emph{tentacle function}.
\end{mydef}
Given a graph $G$, we denote with $\nodes G$, $\edges G$, and
$\tentacle G$ its nodes, edges, and tentacle function,
respectively.
An edge $e \in \edges G$ is connected to a list of nodes via
$\tentacle G$ 
and the \emph{arity} of $e$ is the length of $t_G(e)$.

\begin{mydef}[Graph morphism]\label{morphismDef}
  Let $G$ and $H$ be two graphs.  A \emph{graph morphism from $G$ to
    $H$} is a pair of functions $\anglebraces{ \sigma_V:\nodes G \to
    \nodes H, \sigma_E :\edges G \to \edges H}$ s.t. $\sigma_V$ and
  $\sigma_E$ preserve the tentacle functions, i.e.
  $\after{\sigma_V^*}{\tentacle G} = \after{\tentacle H}{\sigma_E}$,
  where $\sigma_V^*$ is the homomorphic extension of $\sigma_V$ to $\nodes
  G^*$.
\end{mydef}

In ADR, graphs are typed over a fixed type graph via typing morphisms.
A graph \emph{$G$ is typed over a type graph $\tg$ through $\typing G$} if
$\typing G$ is a morphism from $G$ to $\tg$.
\begin{mydef}[ADR graph]
  Let $\tg$ be a type graph equipped with a map 
  $\eta : \edges \tg \to \{0,1\}$.
  An ADR graph \emph{$G$ is a (hyper)graph typed over $\tg$
    through $\typing G$} if
  $\typing G$ is a morphism from $G$ to $\tg$; we call $e \in \edges G$
  \emph{terminal} if $\eta(\sigma(e)) = 0$ and \emph{non-terminal} if
  $\eta(\sigma(e)) = 1$.
\end{mydef}
This is reminiscent of string grammars where terminal symbols
correspond to terminal edges and non-terminal symbols to non-terminal
edges.

\begin{example}\label{ex:tgg}
  Let $V = \set{\chainport} \subseteq \nodeset$ and $E=
  \set{\client, \bookF, \findF, \flights, \flight, \pay, \fpay}
  \subseteq \edgeset$.
  Consider the type graph $\tg = \anglebraces{V, E, t,\eta}$ where
  $\tentfun t \client {\chainport}$ and $\tentfun t e
  {\chainport , \chainport} $ for each $e \in E \setminus \set\client$,
  with $\eta(e) = 0$ if $e \in \set{\client,\findF}$ and
  $\eta(e) = 1$ otherwise.
  The graph $G = \anglebraces{\set{u_1, \dots ,u_4},\set{ff,fl_1,fl_2},
    t'}$ where $t'$ is defined as $\tentfun{t'}{ff}{u_2,u_1}$,
  $\tentfun{t'}{fl_1}{u_3,u_2}$, and $\tentfun{t'}{fl_2}{u_4, u_2}$ can
  be typed on $\tg$ by $\typing G$ mapping all the nodes to $\bullet$,
  $fl_1$ and $fl_2$ to $\flights$, and $ff$ to $\findF$.
\end{example}

Hereafter, we fix a typed graph $\tg$ and tacitly assume that all
graphs $G$ are typed over $\tg$ via a morphism $\typing G$.
Intuitively, $\tg$ yields the \emph{vocabulary} of the architectural
elements to be used in the designs; moreover, $\tg$ specifies how
these elements can be connected together (e.g., as in
Example~\ref{ex:tgg}).

Type and typed graphs have a convenient visual notation.
Nodes are circles and edges are drawn as (labelled) boxes;
single- and double-lined boxes represent terminal and non-terminal
edges, respectively.
Tentacles are depicted as lines connecting boxes to circles;
conventionally, directed tentacles indicate the first node attached to
the edge and the others are taken clockwise.
The visual notation for typed graphs include the graph and its typing
morphism.
Nodes are paired with their types while an edge label $e : e'$
represents the fact that the typing morphism maps the edge $e$ of the
graph to the edge $e'$ of the type graph.
\begin{example}\label{ex:tgandg}
  In the visual notation described above, the type graph
  $\tg$ and the graph $G$ of Example~\ref{ex:tgg} can be respectively
  drawn as
  {\small\[
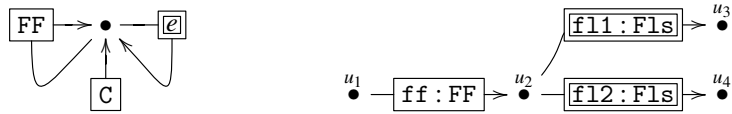
  \begin{array}{c@{\hspace{2cm}}c}
    \xymatrix@C=.5cm@R=.5cm{
      \tedge{\findF} \ar@(r,l)[r] \ar@{-}@(d,dl)[r]
      & {\chainport}
      & \ntedge{e} \ar@(d,dr)[l] \ar@{-}@(l,r)[l]
      \\
      & \tedge{\client} \ar@(u,d)[u]
    }
    &
    \xymatrix@C=.3cm@R=.3cm{
      & & &
      \ntedge{\flightsobj{\texttt{fl1}}} \ar@(r,l)[r] \ar@{-}@(l,ur)[dl]
      &\stackrel{u_3}{\chainport}\\
      \stackrel{u_1}{\chainport} &
      \tedge{\findFobj{\texttt{ff}}} \ar@(r,l)[r] \ar@{-}@(l,r)[l]
      & \stackrel{u_2}{\chainport} &
      \ntedge{\flightsobj{\texttt{fl2}}} \ar@(r,l)[r] \ar@{-}@(l,r)[l]
      & \stackrel{u_4}{\chainport} &
    }
  \end{array}\]}
where, to simplify the type graph, we use $e \in \{\bookF, \flights,
\flight, \pay, \fpay\}$ (instead on drawing an edge for each
non-terminal edge of $\Gamma$.
\end{example}

\begin{mydef}[Typed Graph morphisms]\label{tgDef}
  A \emph{morphism} between $\tg$-typed graphs $f: G_1 \to G_2$ is a
  \emph{typed graph morphism} if it preserves the typing, i.e. such
  that $\typing{G_1} = \typing{G_2} \circ f$.
\end{mydef}

\begin{mydef}[Productions]\label{dpDef}
  A \emph{(design) production} $p$ is a tuple $\anglebraces{ L,R,
    i:\nodes L \to \nodes R}$ where $L$ is a graph consisting only of
  a non-terminal edge attached to distinct nodes; $R$ is an ADR graph
  (with both terminal and non-terminal edges); the nodes in $Im(i)$
  (the image of $i$) are called \emph{interface nodes}.
\end{mydef}

Design productions can be thought of as rewriting rules that, when
applied to a graph $G$, replace 
a non-terminal (hyper)edge of $G$ matching $L$
 with a fresh copy of
$R$ (we remark that
our morphisms are type-preserving). Also productions have a
suitable visual representation illustrated in the next example.
\begin{example}\label{ex:chainProds}
  The graphical representation below represents a design production.
  \[ \begin{array}{c@{\hspace{4cm}}c}
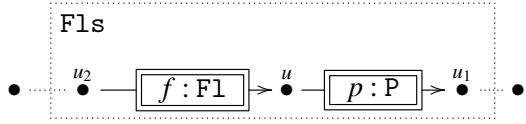
    \begin{minipage}{8cm}{$
        \xymatrix@C=.3cm@R=.3cm{
          &
          \flights
          \\
          \chainport \ar@{.}[r] & \stackrel{u_2}{\chainport} &
          \ntaritytwo{\ \ \flightobj{f} \ \ }{r}{l} &
          \stackrel{u}{\chainport} &
          \ntaritytwo{\ \ \payobj{p}\ \ }{r}{l}&
          \stackrel{u_1}{\chainport} & \chainport \ar@{.}[l]
          \save "1,2"."2,6"*+[F.] \frm{}
          \restore 
        }
      $}
    \end{minipage}
  \end{array}\]
  Since the production above will be used later
  (cf. Example~\ref{ex:path}) we will refer to it as
  $\bookflightprod$.  The left-hand-side (LHS) of $\chainprod$ is an
  edge of type $\flights$ (denoted in the left-upper corner of the
  dotted-box) whose nodes are those outside the dotted box; we omit
  the identities of such nodes when immaterial. The right-hand-side
  (RHS) of $\bookflightprod$ is the graph inside the dotted box.
  The mapping $i$ of $\bookflightprod$ is represented by the dotted lines.
\end{example}

The application of asserted productions
(cf. Definition~\ref{def:application}) encompasses that of
ADR productions hence we give here only an example to illustrate how
productions are applied.

\begin{example}\label{ex:chainApplic}
  Consider the production $\chainprod$ of Example~\ref{ex:chainProds}.
  In the following rewriting
  {\[\begin{array}{c@{\hspace{2cm}}c}
    \def\g#1{\save
      [].[dr]!C="g#1"\frm{}\restore}%
    
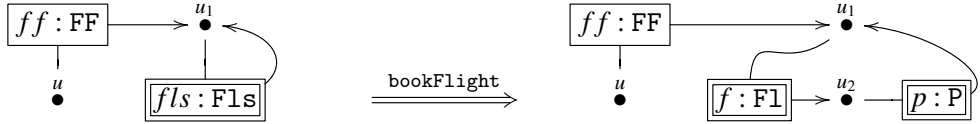
\xymatrix@C=.5cm@R=.3cm{
      \tedge{\findFobj{ff}}
      \ar@(r,l)[r] \ar@(d,u)@{-}[d]
      &\stackrel{u_1}{\chainport}&&&&&
      \g2&
      \tedge{\findFobj{ff}}
      \ar@(r,l)[rr] \ar@(d,u)@{-}[d]
      &&\stackrel{u_1}{\chainport}
      &   \\
      \stackrel{u}{\chainport}
      &\ntedge{\flightsobj{fls}}
      \ar@(r,r)[u] \ar@(u,d)@{-}[u]
      &&&&&
      & \stackrel{u}{\chainport}
      & \ntedge{\flightobj{f}}
      \ar@(r,l)[r] \ar@(u,dl)@{-}[ur]
      &\stackrel{u_2}{\chainport}
      & \ntedge{\payobj{p}}
      \ar@(r,r)[ul] \ar@(l,r)@{-}[l]
      \ar @{=>} "2,4" ;"2,7" ^-{\bookflightprod}
    }
    \end{array}\]}
  the unique edge of type $\flights$ in the leftmost graph is replaced
  by an instance of the RHS of $\bookflightprod$.
  Note that the rest of the graph (consisting only of the edge $ff$)
  including the interface nodes is left unchanged while a fresh
  node $u_2$ is created.
\end{example}


\section{A logic for ADR}
\label{sec:logic}
We use a simple logic tailored on ADR.
Basically, our logic is a propositional logic to predicate on
(in)equalities of nodes.
In the following we let $D, D', \ldots$ range over edges of $\tg$.
\begin{mydef}[ADR logic]\label{def:logic}
  Let $\varset$ be a countably infinite set of variables for nodes
  (ranged over by $\lvar x, \lvar y, \lvar z, \ldots$).
  The set $\L$ of \emph{(graph) formulae} for ADR is given by the
  following grammar:
  \[
  \psi, \varphi ::=  \quad \lvar x = \lvar y
  \quad | \quad      \true
  \quad | \quad      \neg \varphi
  \quad | \quad      \varphi_1 \land \varphi_2   
  \quad | \quad      \forall D(\tilde{\lvar x}).\varphi
  \]
  In formulae of the form $\forall D(\tilde{\lvar x}).\varphi$, the
  occurrences of $\lvar y \in \tilde{\lvar x}$ in $\varphi$ are
  \emph{bound}, $\tilde{\lvar x}$ has the length of the arity of $D$
  and $\tilde{\lvar x}$ are pairwise distinct.
\end{mydef}
Logic $\L$ is parametrised with respect to the type graph $\tg$ used
in quantification.
Variables not in the scope of a quantifier are free and the set $\fv
\varphi$ of \emph{free variables} of $\varphi \in \L$ is defined
accordingly; also, we abbreviate $\lvar x_1 = \lvar x_2 \land \ldots
\land \lvar x_{n-1} = \lvar x_n$ with $\lvar x_1 = \lvar x_2 = \ldots
= \lvar x_{n-1} = \lvar x_n$ and we define $\bot$ as $\neg \top$,
$\lvar x \neq \lvar y$ as $\neg (\lvar x = \lvar y)$, $\varphi \vee
\psi$ as $\neg ( \neg \varphi \land \neg \psi)$,  $\varphi \to \psi$
as $\neg \varphi \vee \psi$ , and $\exists D(\tilde{\lvar x}).\varphi$
as $\neg \forall D(\tilde{\lvar x}).\neg \varphi$.

\bigskip

The models of our logical formulae are ADR graphs.
\begin{mydef}[Satisfaction relation]\label{def:entailment}
  An ADR graph $G$ \emph{satisfies $\varphi \in \L$ under the
    assignment $h : \varset \to \nodes G$} (in symbols $G
  \models_h \varphi$) iff
  \[\begin{array}{lcll}
    \varphi \equiv \true,
    & & & or 
    \\
    \varphi \equiv \lvar x = \lvar y
    & \text{and} &
    h(\lvar x) = h(\lvar y),
    & or
    \\
    \varphi \equiv \neg \varphi'
    & \text{and} &
    G \nvDash_h \varphi' ,
    & or
    \\
    \varphi \equiv \varphi_1 \land \varphi_2 
    & \text{and} &
    G \models_h \varphi_1 \text{ and } G \models_h \varphi_2 ,
    & or
    \\
    \varphi \equiv \forall D(\tilde{\lvar x}).\varphi
    & \text{and} &
    G \models_{h \tsubs {\lvar x} u} \varphi
    & \text{ for any } d(\tilde u) \in G  \text{ s.t. } \tau_G(d)= D
  \end{array}\]
\end{mydef}
Note that in the last case of Definition~\ref{def:entailment}, each bound
variable in $\tilde{\lvar x})$ is replaced with a node.

\noindent
\textbf{Fact.\ } For each $h,h' : \varset \to \nodes G$, if $h|_{\fv \varphi} = h'|_{\fv \varphi}$
then $G \models_h \varphi$ iff  $G \models_{h'} \varphi$.

\noindent
By the above property, in $G \models_h \varphi$ we can restrict to
finite mappings $h$ that only assign variables in $\fv \varphi$.
Hereafter, we write $G \models \varphi$ when $\fv \varphi = \emptyset$.

\begin{example}\label{ex:test}
  The formula $\phi_{\text{ex}} = \forall D(\lvar x, \lvar y).\exists D'(\lvar z). \lvar
  x = \lvar z$ describes graphs such that each edge of type $D$ is
  connected to one of type $D'$ on the first tentacle.
  For instance, consider the graphs
  {\small\[
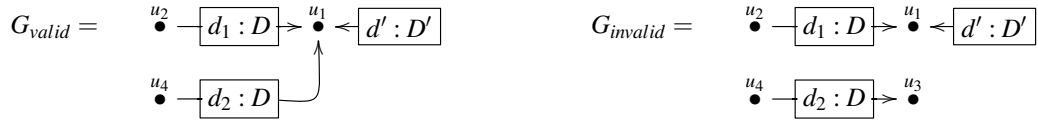
    \begin{array}{c@{\hspace{2cm}}c}
        G_{valid} =
      \xymatrix@C=.3cm@R=.3cm{
        &\stackrel{u_2}{\chainport}
        &\tedge{d_1:D} \ar@(r,l)[r] \ar@{-}@(l,r)[l]
        &\stackrel{u_1}{\chainport}
        &\tedge{d':D'} \ar@(l,r)[l]\\
        &\stackrel{u_4}{\chainport} 
        &\tedge{d_2:D} \ar@(r,d)[ur] \ar@{-}@(l,r)[l]
      }
      &
      G_{invalid}  =
      \xymatrix@C=.3cm@R=.3cm{
        &\stackrel{u_2}{\chainport}
        &\tedge{d_1:D} \ar@(r,l)[r] \ar@{-}@(l,r)[l]
        &\stackrel{u_1}{\chainport}
        &\tedge{d':D'} \ar@(l,r)[l]\\
        &\stackrel{u_4}{\chainport} 
        &\tedge{d_2:D} \ar@(r,l)[r] \ar@{-}@(l,r)[l]
        &\stackrel{u_3}{\chainport} 
      }
    \end{array}\]}
  then $G_{valid}$ satisfies $\phi_{\text{ex}}$ whereas $G_{invalid}$
  does not because $d_2$ is not connected to any edge of type $D'$.
\end{example}

More interesting formulae are given in the next two examples.
\begin{example}\label{ex:interesting}
  The formula
  \begin{equation}\label{eq:notexists}
    \no D \mmdef \forall D(\tilde{\lvar x}).\false    
  \end{equation}
  characterises the graphs that do not contain edges of a given type.
\end{example}
Formulae of the form~\eqref{eq:notexists} will be used in
Definition~\ref{def:wp} (hereafter, we write $\no {D_1,\ldots,D_n}$
for $\no{D_1} \land \ldots \land \no{D_n}$).

The next example shows that, despite its simplicity, our logic is
quite expressive when ``taken modulo productions''.

\begin{example}\label{ex:path}
  By the production below, a non-terminal edge of type $C$ can be
  replaced by a chain of two edges of type $C$.
  The formula $\pa D{\classes}$ requires instead that any two different
  nodes attached to an edge of type $D$ are connected by an edge of type $C$.
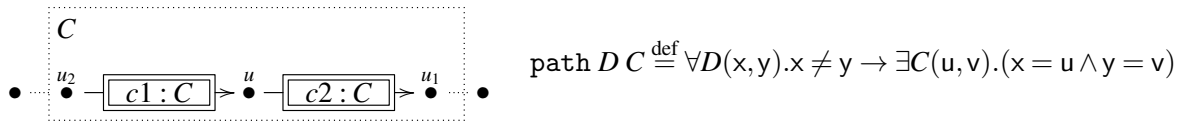
  \[ \begin{array}{cc}
    \begin{minipage}{6.7cm}{$
        \xymatrix@C=.25cm@R=.25cm{
          &
          C
          \\
          \chainport \ar@{.}[r] & \stackrel{u_2}{\chainport} &
          \ntaritytwo{\ \ {c1:C} \ \ }{r}{l} &
          \stackrel{u}{\chainport} &
          \ntaritytwo{\ \ {c2:C}\ \ }{r}{l}&
          \stackrel{u_1}{\chainport} & \chainport \ar@{.}[l]
          \save "1,2"."2,6"*+[F.] \frm{}
          \restore 
        }
        $}
    \end{minipage}
    & \pa D{\classes} \mmdef
    \forall D(\lvar x, \lvar y).\lvar x \neq \lvar y \to
    \exists C(\lvar u, \lvar v). (\lvar x = \lvar u \land \lvar y = \lvar v )
  \end{array}\]
  The production and the formula above characterise graphs that
  contain paths of edges of type $C$ between any two distinct nodes
  connected by an edge of type $D$.
  Note that even though there is no edge of type $D$ in the
  production, $\pa D C$ quantifies over edges of type $D$ in the
  graph.
\end{example}


\section{Design by Contract for ADR}
\label{sec:dbc}
Our notion of contracts hinges on \emph{asserted productions}, namely
ADR productions decorated with pre- and post-conditions expressed in
the logic $\L$ given in \S~\ref{sec:logic}.
\begin{mydef}[Asserted productions]
  \label{def:applicability}\label{def:assertedprod}
  If $p = \anglebraces{L,R,i}$ is a production, $h, h': \varset \to
  \nodeset$, and $\psi, \varphi \in \L$ then $\prd{\psi , h} p
  {\varphi , h'}$ is an \emph{asserted production} iff $h(\fv \psi)
  \subseteq \nodes L$, and $h'(\fv \varphi) \subseteq \nodes R$.
\end{mydef}
An asserted production generalises ADR productions and it intuitively
requires that if $p$ is applied to a graph $G$ that satisfies $\psi$
then the resulting graph is expected to satisfy $\varphi$.
The maps $h$ and $h'$ in Definition~\ref{def:assertedprod} allow pre-
and post-conditions to predicate on nodes occurring in the LHS or the
RHS of $p$.

\begin{figure}[t]\centering
  $\xymatrix@R1.5em@C3em{
    \psi \ar[r]^h & L\ar[r]^i\ar[dd]^{\sigma} & R\ar@{=}[d]^\iota & \ar[l]_{h'}\varphi
    \\
    &   & R'\ar@{^{(}->}[d]
    \\
    & G\ar@{|=}[uul]\ar[r]^\pi & G'
  }$
  \caption{\label{fig:dpT}Asserted design productions}
\end{figure}
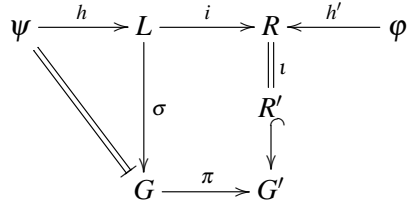

An \emph{instance $G'$ of a graph $G$} is a graph $G'$ isomorphic to
$G$ that does not share nodes or edges with $G$.
The application of an asserted production to a graph consists of
replacing an homomorphic image of the edge of the LHS with a new
instance of the RHS and then connecting it to the interface nodes.
This is formalised in the next definition and schematically
illustrated in Figure~\ref{fig:dpT}.

\begin{mydef}[Applying asserted productions]\label{def:application}
  Let $p = \anglebraces{L,R,i}$ be a production, $G$ a graph,
  and $\sigma$ a morphism from $L$ to $G$.
  We say that $\pi = \prd {\psi,h} p {\varphi,h'}$, an asserted production,
  \emph{is applicable to $G$ via $\sigma$} iff
  $G \models_{\after \sigma h} \psi$.

  Given an instance $R'$ of $R$ through the isomorphism $\iota : R
  \to R'$ such that $\edges{R'} \cap \edges G = \emptyset$ and
  $\nodes{R'} \cap \nodes{G} = \emptyset$ a graph
  $(G' =) G\subs{\sigma(e)}{R''}$ is the \emph{application of $\pi$
    to $G$ wrt $\sigma$} iff $R'' =
  R'\subs{\iota(r)}{\sigma(i^{-1}(r)) \ \st\ r \in Im(i)}$.
  A production $\pi$ is \emph{valid} when any application of $\pi$ to
  a graph satisfying the precondition of $\pi$ yields a graph
  satisfying the post condition of $\pi$.
\end{mydef}

Examples~\ref{ex:applicability} and~\ref{ex:application} show how
asserted productions are applied to graphs.

\begin{example}\label{ex:applicability}
  Let $\psi \mmdef \forall \flights(\lvar x, \lvar y).\lvar x \neq \lvar y$
  and let $\pi \mmdef \prd{\psi,\emptyset}
  \bookflightprod {\phi,\emptyset}$ be an asserted production of
  $\bookflightprod$ given in Example~\ref{ex:chainProds}.
  If $G$ is the leftmost graph in the rewriting of
  Example~\ref{ex:chainApplic}, then
  we have $G \not\models \psi$ (under the unique morphism $\sigma$ from
  $L$ to $G$).
  In fact, $\lvar x$ and $\lvar y$ are mapped to the same node
  $u_1$ of $G$.
\end{example}
\begin{example}\label{ex:application}
  The rewriting below
{\[\begin{array}{c@{\hspace{1.75cm}}c}
    \def\g#1{\save
      [].[dr]!C="g#1"\frm{}\restore}%
    
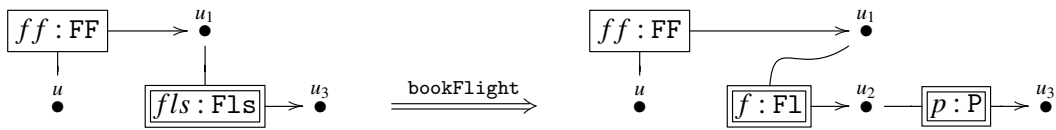
\xymatrix@C=.5cm@R=.3cm{
      \tedge{\findFobj{ff}}
      \ar@(r,l)[r] \ar@(d,u)@{-}[d]
      &\stackrel{u_1}{\chainport}&&&&&
      \g2&
      \tedge{\findFobj{ff}}
      \ar@(r,l)[rr] \ar@(d,u)@{-}[d]
      &&\stackrel{u_1}{\chainport}
      &   \\
      \stackrel{u}{\chainport}
      &\ntedge{\flightsobj{fls}}
      \ar@(r,l)[r] \ar@(u,d)@{-}[u]
      & \stackrel{u_3}{\chainport}
      &&&&
      & \stackrel{u}{\chainport}
      & \ntedge{\flightobj{f}}
      \ar@(r,l)[r] \ar@(u,dl)@{-}[ur]
      &\stackrel{u_2}{\chainport}
      & \ntedge{\payobj{p}}
      \ar@(r,l)[r] \ar@(l,r)@{-}[l]
      & \stackrel{u_3}{\chainport}
      \ar @{=>} "2,4" ;"2,7" ^-{\bookflightprod}
    }
  \end{array}\]
}
is obtained by the asserted production $\pi$ in
Example~\ref{ex:applicability};
according to Definition~\ref{def:application}, edge $fls$ on the left
is replaced by an isomorphic instance of $R$ preserving
the interface nodes $u_1$ and $u_3$.
\end{example}
We remark that Definition~\ref{def:application} generalises the
rewriting mechanism (hyper-edge replacement)~\cite{DBLP:conf/gg/DrewesKH97}
of ADR, in fact $\prd{\true,\emptyset} p{\true,\emptyset}$
applies exactly as normal ADR productions.


\section{Extracting contracts for ADR productions}
\label{sec:wpc}
The application of an asserted production $\prd{\psi,h} p {\varphi,h'}$
to a graph satisfying $\psi$ does not necessarily yield a graph
satisfying $\varphi$ (this can be trivially noted by taking a
production with $\false$ as post-condition).
We give an algorithm to compute the weakest pre-condition given a
post-condition and a production in the style of the seminal work on
predicate transformers of Dijkstra~\cite{dij75}.
We first give some auxiliary definitions and notations.

Hereafter, bound variables in a formula are assumed distinct from its
free variables and bound only once.
An \emph{environment} $\envirWp$ is the product of three finite partial
maps $\envirWp^{(1)}: \varset \to \set{\forall,\exists}$,
$\envirWp^{(2)}: \varset \to \edges \tg$, and $\envirWp^{(3)}: \varset
\to \nodeset$.
Hereafter, we write $\envempty$ for the empty environment,
$\envirNot{\lvar x} D G q$ when $\lvar x$ is quantified by $q \in
\set{\forall,\exists}$ (that is $\envirWp^{(1)}(\lvar x) = q$),
attached to an edge of type $D$ (that is $\envirWp^{(2)}(\lvar x) =
D$), and mapped to node of $G$ (that is $\envirWp^{(3)}(\lvar x) \in
\nodes G$); if $G$ consists of a node $n$, we simply write
$\envirNot{\lvar x} D n q$.
Also, we use $" \_ "$ as a wild-card writing e.g. $\envirNot {\lvar x}
\_ G q$ when we are not interested in the type assigned to $\lvar x$
(i.e., $\envirNot {\lvar x} \_ G q$ abbreviates $\envir{\lvar x}{1} =
q$ and $\envir{\lvar x}{3} \in \nodes G$).

\begin{mydef}[Auxiliary Mapping]\label{def:aux}
Let $p = \anglebraces{L, R, i}$ be a production. We write $\inodes R
\mmdef \nodes R \setminus Im(i)$ to denote the \emph{internal} nodes of
$p$, and $\bar R \mmdef \nodeset \setminus \nodes R$ 
to denote the nodes \emph{outside} $p$.
Given $\psi_1 , \psi_2 , \psi_3 \in \L$ the map
$\fwp p { \psi_1, \psi_2, \psi_3} {\lvar x_1 = \lvar x_2}{\envirWp}$ is:
\[\begin{array}{rll}
  \fwp p {\psi_1,\psi_2,\psi_3} {\lvar x_1 = \lvar x_2} \envirWp & = &
  \begin{cases}
    \true & \text{if}\ \envirNot{\lvar x_1}{\_}{n}{\exists},
    \envirNot{\lvar x_2}{\_}{n}{\exists} \ \text{and}\
    n \in \inodes R
    \\
    \false & \text{if}\ \envirNot{\lvar x_1}{\_}{\inodes R}{\forall} \
    \text{and}\ (\envirNot{\lvar x_2}{\_}{\bar{R}}{\exists} \ \text{or} \
    \envirNot{\lvar x_2}{\_}{Im(i)}{\exists})
    \\
    \false & \text{if}\ \envirNot{\lvar x_1}{\_}{\inodes R}{\forall},
    \envir{\lvar x_2} 3 \in \inodes R \ \text{and}\ \envir {\lvar x_1} 3 \neq \envir {\lvar x_2} 3
    \\
    \psi_1 & \text{if}\ \envirNot{\lvar x_1}{\_}{\inodes R}{\forall} \
    \text{and}\ \envirNot{\lvar x_2}{D}{\bar{R}}{\forall}
    \\
    \psi_2 & \text{if}\ \envirNot{\lvar x_1}{D}{n}{\forall} \
    \text{and}\ \envirNot{\lvar x_2}{D'}{n}{\forall} \ \text{and}\
    n \in \inodes R
    \\
    \psi_3 & \text{otherwise}
  \end{cases}
\end{array}
\]
that, depending on $\envirWp$, returns either $\psi_j$, $\true$, or
$\false$.
\end{mydef}
The map $\fwp p { \psi_1, \psi_2, \psi_3} {\lvar x_1 = \lvar x_2}{\envirWp}$
in Definition~\ref{def:aux} is parametrised with 
$\psi_1$, $\psi_2$, and $\psi_3$.
Intuitively, \\
$\fwp p { \psi_1, \psi_2, \psi_3} {\lvar x_1 = \lvar x_2}{\envirWp}$
inspects the environment $\envirWp$ and returns $\true$, $\false$,
$\psi_1$, $\psi_2$, or $\psi_3$.
The variables $\lvar x_1$ and $\lvar x_2$ in an equality are
quantified/assigned in $\envirWp$. More precisely,
\begin{itemize}
\item $\fwp p { \psi_1, \psi_2, \psi_3} {\lvar x_1 = \lvar x_2}{\envirWp}$
  returns $\true$ when $\lvar x_1$ and $\lvar x_2$ are both existentially
  quantified and assigned to internal nodes of $R$, the RHS of $p$, then
  the application of $p$ guarantees the equality $\lvar x_1 = \lvar
  x_2$ regardless the graph it is applied to;
\item $\fwp p { \psi_1, \psi_2, \psi_3} {\lvar x_1 = \lvar x_2}{\envirWp}$
  returns $\false$ when one of the nodes, say $\lvar x_1$ is universally
  quantified and assigned to an internal node of $R$ while $\lvar x_2$
  is either not internal or internal but assigned to a different node
  than $\lvar x_1$;
\item in the other cases,
  $\fwp p { \psi_1, \psi_2, \psi_3} {\lvar x_1 = \lvar x_2}{\envirWp}$
  returns either $\psi_1$, $\psi_2$, or $\psi_3$; as it will be more
  clear after Definition~\ref{def:wd}, such conditions state the
  absence of some edges from the graph $p$ is applied to or the
  validity of a suitable node equality.
\end{itemize}

A formula $\phi \in \L$ is in \emph{negation normal formal form} when
it is closed and negation occurs only in front of equalities.
It is trivial to see that all formulae of $\L$ have an equivalent
negation normal form.

\begin{mydef}[Weakest pre-conditions] \label{def:wp}\label{def:wd}
  Let $p = \anglebraces{L, R, i}$ be a production, $\envirWp$ an
  environment and $\mathtt Z = \set{\lvar z_1, \ldots, \lvar z_m}
  \subseteq \varset$ where $m$ is the arity of $L$,
  $\varphi \in \L$ in negation normal form, $h :
  \fv \varphi \to \nodes R$ be injective, and $\bar h : \mathtt Z \to
  \nodes L$ a bijection.

  The predicate $\wpre {\bar h} h p \varphi \mmdef \wDeff {p, \psi} {\envirWp}{(\varphi)}
  \land \wp {p,\bar h} {h, \envirWp} \varphi$ ---~where the predicate
  transformers $\wDeff {p, \psi} {\envirWp}{(\varphi)}$ and $\wp {p,\bar h}
  {h, \envirWp} \varphi$ are defined below~--- is the \emph{weakest
    pre-condition of $p$ with post-condition $\varphi$ under $h$,
    $\bar h$}.

  The maps $\wDef {p,\psi} {\envirWp} \varphi$ and $\wp {p,\bar h}
  {h,\envirWp} \varphi$ are defined below where, in the clauses for
  quantifiers $\forall D (\tilde{\lvar x}).\_$ and $\exists D
  (\tilde{\lvar x}).\_$ we assume that $\set{v_1,
    \ldots, v_n} \subseteq \bar R$ is a fixed set of
  (representative) external nodes.
  Also, the condition $\choose u R D$ holds iff $\tilde u \cap \inodes R
  = \emptyset$ when $R$ does not have edges of type $D$.

  \[\begin{array}{l@{\quad}l@{\quad}l}
    \wDef {p,\psi} {\envirWp} {\lvar x_1 = \lvar x_2} & = & \fwp p 
    { \no D, \no{D,D'}, \psi } {\lvar x_1 = \lvar x_2} \envirWp
    \\[.2cm]
    \wDef {p,\psi} {\envirWp} {\lvar x_1 \neq \lvar x_2} & = & 
    \neg \fwp p { \false, \true, \neg \psi}  {\lvar x_1 = \lvar x_2} \envirWp
    \\[.2cm]
    \wDef {p,\psi} {\envirWp} \true & = & \true
    \\[.2cm]
    \wDef {p,\psi} {\envirWp} {\phi \land \phi'} & = & \wDef {p,\psi} {\envirWp} \phi \land \wDef {p,\psi} {\envirWp} {\phi'}
    \\[.2cm]
    \wDef {p,\psi} {\envirWp} {\phi \lor \phi'} & = & \wDef {p,\psi} {\envirWp} \phi \lor \wDef {p,\psi} {\envirWp} {\phi'}
    \\[.2cm]
    \wDef {p,\psi} {\envirWp} {\forall D (\tilde{\lvar x}).\phi} & = &
    \displaystyle{\bigwedge_{\choose u R D}} \wDef {p,\psi} {\envirWp'}
    \phi
    \\ & 
    \text{where} & \tilde{\lvar x} = \lvar x_1,
    \ldots, \lvar x_n \text{ and } \tilde u = u_1, \ldots, u_n \in (\nodes R
    \cup \set{v_1, \ldots, v_n})^*
    \\ &
    \text{and} & \envirWp' =
    \envirWp\subs{\lvar x_j}{(\forall,D,u_j) \ \st\ j=1,\ldots,n}
    \\[.2cm]
    \wDef {p,\psi} {\envirWp} {\exists D (\tilde{\lvar x}).\phi} & = &
    \displaystyle{\bigvee_{\choose u R D}} \wDef {p,\psi} {\envirWp'}
    \phi
    \\ & 
    \text{where} & \tilde{\lvar x} = \lvar x_1,
    \ldots, \lvar x_n \text{ and } \tilde u = u_1, \ldots, u_n \in (\nodes R
    \cup \set{v_1, \ldots, v_n})^*
    \\ &
    \text{and} & \envirWp' =
    \envirWp\subs{\lvar x_j}{(\exists,D,u_j) \ \st\ j=1,\ldots,n}
  \end{array}\]

  %
  \[\begin{array}{l@{\quad}l@{\quad}l}
    \wp {p,\bar h} {h, \envirWp} {\lvar x_1 = \lvar x_2} & = & 
    \fwp p { \no{D}, \no{D,D'}, \lvar y_1 = \lvar y_2 } 
    {\lvar x_1 = \lvar x_2} \envirWp
    \\
    &
    \text{where } &  \lvar y_j = \bar h^{-1}(i^{-1}(h(\lvar x_j))) 
    \text{ if } h(\lvar x_j) \in Im(i) 
    \text{, and } \lvar y_j = \lvar x_j \text{ otw}
    \\[.2cm]
    \wp {p,\bar h} {h, \envirWp} { \lvar x_1 \neq  \lvar x_2 } & = & 
    \neg \fwp p { \lvar y_1 = \lvar y_2, \true, \lvar y_1 = \lvar y_2 } 
    {\lvar x_1 = \lvar x_2} \envirWp
    \\&
    \text{where } & \lvar y_j = \bar h^{-1}(i^{-1}(h(\lvar x_j))) 
    \text{ if } h(\lvar x_j) \in Im(i) \text{, and } \lvar y_j = \lvar x_j \text{ otw}
    \\[.2cm]
    \wp {p,\bar h} {h, \envirWp} \true & = & \true
    \\[.2cm]
    \wp {p,\bar h} {h, \envirWp} {\phi \land \phi'} & = & \wp {p,\bar h} {h, \envirWp} \phi \land \wp {p,\bar h} {h, \envirWp} {\phi'}
    \\[.2cm]
    \wp {p,\bar h} {h, \envirWp} {\phi \lor \phi'} & = & \wp {p,\bar h} {h, \envirWp} \phi \lor \wp {p,\bar h} {h, \envirWp} {\phi'}
    \\[.2cm]
    \wp {p,\bar h} {h, \envirWp} {\forall D (\tilde{\lvar x}).\phi}& = &
    \displaystyle{\bigwedge_{\choose u R D}} \forall D (\tilde{\lvar x}) .
    \wp {p,\bar h} {h, \envirWp'} \phi
    \\ & \text{where} &
    \tilde{\lvar x} = \lvar x_1,
    \ldots, \lvar x_n \text{ and } \tilde u = u_1, \ldots, u_n \in (\nodes R
    \cup \set{v_1, \ldots, v_n})^*
    \\ & \text{and} &
    \envirWp' =
    \envirWp\subs{\lvar x_j}{(\forall,D,u_j) \ \st\ j=1,\ldots,n}
    \\[.2cm]
    \wp {p,\bar h} {h, \envirWp} {\exists D (\tilde{\lvar x}).\phi}& = &
    \displaystyle{\bigvee_{\choose u R D}} \big( \exists D (\tilde{\lvar x}) .
    \wp {p,\bar h} {h, \envirWp'} \phi 
    \lor  \wDef {p,\false} {h, \envirWp'} \phi 
    \big)
    \\ & \text{where} &
    \tilde{\lvar x} = \lvar x_1,
    \ldots, \lvar x_n \text{ and } \tilde u = u_1, \ldots, u_n \in (\nodes R
    \cup \set{v_1, \ldots, v_n})^*
    \\ & \text{and} &
    \envirWp' =
    \envirWp\subs{\lvar x_j}{(\exists,D,u_j) \ \st\ j=1,\ldots,n}
  \end{array}
  \]
\end{mydef}
The weakest pre-condition is the conjunction of the predicates
computed by the predicate transformers $\wDeff {p, \psi} {\envirWp}$ and 
$\wpf{p,\bar h}{h, \envirWp}$ on the post condition $\varphi$.
The first transformer simply checks that the production $p$ can
guarantee the post-condition for some pre-condition.

The most interesting cases in Definition~\ref{def:wp} are the ones for
equality $\lvar x_1 = \lvar x_2$ dealt by the auxiliary map $\fwp p
{\psi_1, \psi_2, \psi_3} {\lvar x_1 = \lvar x_2}{\envirWp}$.
If both $\lvar x_1$ and $\lvar x_2$ are existentially quantified and
assigned to the same internal nodes of $p$, the calculated weakest
pre-condition is $\true$; in fact, whatever graph the production is
applied to, the post-condition would be guaranteed by the RHS of $p$.
Instead $\false$ is returned when say $\lvar x_1$ is universally
quantified and ($i$) $\lvar x_2$ is assigned to an interface node and
it is existentially quantified variable, or ($ii$) it is assigned to
an internal node of $R$ different from the one assigned to $\lvar
x_2$.
(Note that in ($i$) if $\lvar x_2$ were universally quantified, there
might be a chance to guarantee the equality if no edges of the type
quantifying the variables were in the graph $p$ is applied to.)
In fact, $\fwp p {\psi_1, \psi_2,\psi_3} {\lvar x_1 = \lvar x_2}
\envirWp$ returns $\false$ if ($i$) $\lvar x_1$ is mapped to a fresh
node in the RHS of $p$ (i.e., an internal node of $p$) while $\lvar
x_2$ is mapped to a node outside $p$ or ($ii$) if they are mapped to
two fresh nodes of the RHS of $p$ because the semantics of ADR does
not allow such identifications on the internal nodes of a production.
The equality $\lvar x_1 = \lvar x_2$ may hold if $\lvar x_1$ and
$\lvar x_2$ are mapped on the same internal node provided that no edge
in the graph $p$ is applied to is typed as the type of the edges
insisting on the variables, otherwise the universal quantification
will be spoiled.
Likewise, if both variables are universally quantified but one is
internal and the other is external (not in $p$), then the weakest
pre-condition returns $\no D$ where $D$ is the type of the external
variable.
Intuitively, the graph resulting from the application of $p$ to a
graph with an $e$ edge of type $D$, would violate the
quantification of $\lvar x_1$ and $\lvar x_2$ since $e$ cannot
insist on fresh nodes introduced by $p$.
In all other cases, $\wp {p,\bar h} {h, \envirWp} {\lvar x_1 = \lvar x_2}$
requires the initial graph to satisfy the
same equality on the nodes corresponding to the variables of the
post-condition; this requires that if either $\lvar x_1$ and $\lvar
x_2$ are assigned to an interface node (that is $h(\lvar x_j) \in
Im(i)$) it has a counterpart variable $\lvar z \in \set{\lvar z_1,
  \ldots, \lvar z_m}$ mapped (through $\bar h$) on the node $i^{-1}(\lvar
x_1)$ or $i^{-1}(\lvar x_2)$ in $L$.

The remaining cases are trivial but for the quantifications $\forall
D(\tilde{\lvar x}).\phi$ and $\exists D(\tilde{\lvar x}).\phi$ 
where the computed pre-conditions require $\phi$ to be satisfied under
any ``reasonable'' assignment to $\tilde{\lvar x}$ for the universal
quantification or one ``reasonable'' assignment to $\tilde{\lvar x}$
for the existential quantification; this means that
such variables are assigned in any possible way either to nodes in
$R$ or to a fixed set of nodes $v_1, \ldots, v_n$ outside $R$;
the choice of such nodes is immaterial the crucial point being just that
they refer to nodes outside $R$ (i.e., as many as the variables in
$\tilde{\lvar x}$).

\begin{proposition}
  If $\psi$ and $\varphi$ are logically equivalent $\L$-formulae, then
  $\wDef {p,\psi} {\envirWp}{\psi}$ (resp. $\wp {p,\bar h} {h,\envirWp} \psi$) is
  logically equivalent to $\wDef {p,\psi} {\envirWp}{\varphi}$ (resp. $\wp
  {p,\bar h}{h,\envirWp} \varphi$).
\end{proposition}

The next example shows how to compute weakest pre-conditions.

\begin{example}\label{ex:wpexample}
  Consider $\varphi \in \L$ and the production $p$ below; let $R$
  be the RHS of $p$:
  \[ \begin{array}{c@{\hspace{1.7cm}}c}
    \varphi \mmdef \forall B(\lvar x, \lvar y). \forall C(\lvar z).\lvar y = \lvar z
    &
    p \mmdef \begin{minipage}{6cm}{$
        \xymatrix@C=.3cm@R=.01cm{
          &{\mathtt{A}}
          \\
          & & & \stackrel{u}{\accessport}
          & \tedge{\mathtt{b:B}} \ar[r] \ar@{-}[l]&
          \stackrel{u_1}{\chainport} && \chainport \ar@{.}[ll]  
          \save "1,2"."2,6"*+[F.] \frm{} \restore
        }$}
    \end{minipage}
  \end{array}\]
  The first step to compute
  $ \wpre {\bar h} h p \phi \mmdef \wDef {p,\true} {\envempty} \varphi \land 
  \wp {p,\bar h} {\emptyset, \envempty} \varphi$ where $\bar h$ refers to the
  interface nodes applies the quantification case in Definition~\ref{def:wp}
  and yields
  \[ 
  \big(\bigwedge_{j=1,2,3}
  \wDef {p,\true} {\envirWp_j} {\varphi'} \big)
  \quad \land \quad
  \big( \bigwedge_{j=1,2,3}
  \forall B(\lvar x, \lvar y).\wp {p,\bar h} {\envempty, \envirWp_j} {\varphi'}
  \ \big)
  \]
  given that $\envirWp_1 = \curlybraces{ \lvar x \mapsto (\forall , B,
    u_1), \lvar y \mapsto (\forall , B, u) }$, $\envirWp_2 =
  \curlybraces{ \lvar x \mapsto (\forall , B, u_1), \lvar y \mapsto
    (\forall , B, v_1) }$ and $\envirWp_3 = \curlybraces{ \lvar x
    \mapsto (\forall , B, v_1), \lvar y \mapsto (\forall , B, v_2) }$
  are the only assignments to consider (since $v_1$ and $v_2$ are
  representative nodes outside $R$ while $u_1$ the unique node on
  $R$'s interface, and $u$ its unique internal node).
  
  The second step applies again this case for $\forall C(\lvar z)$ 
  (for both $\wDef {p,\true} {\envirWp_j} {\varphi'}$ 
  and $\wp {p,\bar h} {\envempty, \envirWp_j} {\varphi'}$) and yields
  \[
  \big( \bigwedge_{j,k=4,5}
  \wDef {p,\true} {\envirWp_j \cup \envirWp_k}{\varphi''}
  \big)
  \quad \land \quad
  \big( \bigwedge_{j,k=4,5}
  \forall B(\lvar x, \lvar y).\forall C(\lvar z).
  \wp {p,\bar h} {\envempty, \envirWp_j \cup \envirWp_k}{\varphi''}
  \big)
  \]
  where $\envirWp_4 = \curlybraces{\lvar z \mapsto (\forall , C, u_1)}$
  and $\envirWp_5 = \curlybraces{\lvar z \mapsto (\forall , C, v_1)}$;
  in fact there is no edge of type $C$ in the RHS of $p$ (hence $v_1$ is
  representative external node and $u_1$ is its unique interface node).

  Finally, applying the auxiliary map $\fwp p {\psi_1, \psi_2,\psi_3}
  {\lvar x_1 = \lvar x_2}{\envirWp}$ for node equality, we get
  \begin{eqnarray}\label{eq:wd}
    \bigwedge_{j,k}
    \wDef {p,\psi} {\envirWp_j \cup \envirWp_k}{\varphi''}
    & \quad = \quad & 
    ( \true \land \no C ) \land (\true \land \true)
    \land (\true \land \true) = \no C
    \\
    \label{eq:wp}
    \bigwedge_{j,k}
    \forall B(\lvar x, \lvar y). \forall C(\lvar z).
    \wp {p,\bar h} {\envempty, \envirWp_j \cup \envirWp_k}{\varphi''}
    & \quad = \quad &
    \forall B(\lvar x, \lvar y). \forall C(\lvar z).\no C
    \; \land \;
    \forall B(\lvar x, \lvar y). \forall C(\lvar z).y=z
  \end{eqnarray}
  Note that, the weakest pre-conditions is the conjunction
  of~\eqref{eq:wd} and~\eqref{eq:wp}, that is 
  \[ \wpre {\bar h} h p \phi \; = \; \no C \; \land \; 
  \forall B(\lvar x, \lvar y). \forall C(\lvar z).\no C
  \; \land \; \forall B(\lvar x, \lvar y). \forall C(\lvar z).y=z
  \]
  this is consistent with the fact that $\phi$ can only be satisfied by graphs that
  do not have any edges of type $C$ due to the internal node $u$
  introduced by the production $p$.
\end{example}

\begin{theorem}
  Let $p = \anglebraces{L, R, i}$ be a production, $\varphi \in \L$,
  $h : \fv \varphi \to \nodes R$ be injective, $\bar h : \mathtt Z \to
  \nodes L$ be a bijection, and $\pi$ be the asserted production
  $\prd{\wpre {\bar h} h p \varphi} p {\varphi, h}$.
  For any ADR graph $G$ and morphism from $L$ to $G$, if $G
  \models_{\after h i} \wpre {\bar h} h p \varphi$ then
    $\pi(G,\sigma) \models_h \varphi$.
\end{theorem}

\begin{theorem}
  For any closed formula $\psi$ such that 
  $\prd{\psi,h'} p {\varphi, h}$ is a valid production then 
  $\psi$ implies $\wpre {\bar h} h p \varphi$.
\end{theorem}


\section{A methodology for recovering invalid configurations}
\label{sec:methodology}
\newenvironment{prodenv}{\begin{minipage}{5cm}\tiny}{\end{minipage}}
In this paper, we envisage architectural styles as formalised by a set
of ADR productions \emph{combined with} a closed formula of our logic
specifying an invariant of the system as illustrated in
Example~\ref{ex:style} below.

\begin{example}\label{ex:style}
  Consider the run-time reconfiguration
  \[\begin{array}{c@{\hspace{2cm}}c}
    \def\g#1{\save
      [].[dr]!C="g#1"\frm{}\restore}%
    
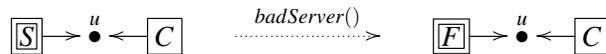
\xymatrix@C=.5cm@R=.3cm{
      \ntarityone{S}{r} 
      & \stackrel{u}{\chainport}
      & \tarityone{C}{l}
      &&&&&
      \ntarityone{F}{r}
      & \stackrel{u}{\chainport}
      & \tarityone{C}{l}
      \ar @{.>} "1,4" ;"1,7" ^-{badServer()}
    }
  \end{array}\]
  where $S$ changes as illustrated to model a failure $F$.
  By imposing an invariant that states that every client has to be
  connected to a non-failed server, the invalid configuration
  can be identified and recovered.
\end{example}

We give a basic methodology for 
recovering a system to a valid state
when run-time configurations compromise it.
We will assume that ADR graphs may be subject to run-time changes.
Instead of giving a formal definition for such graph rewritings, for
the sake of this paper it is enough to consider simple local
rewritings whereby edges may become corrupted and in turn compromise
the desired architectural style in terms of the specified invariant.
In \secref{sec:conc} we briefly
discuss more complex methodologies that we plan to consider in the
future developments.

\bigskip

\newcommand{\inv}{\phi_\text{inv}}

We are interested in computations that start from a system
configuration, say $s_0$, that corresponds to an initial graph,
say $G_0$, supposed to satisfy the invariant, say $\inv$.
The system may evolve at run-time through a series of reconfigurations
($r_i$) that are reflected at the architectural level as schematically
represented in the diagram~\eqref{eq:reconf} below (where $G_i \vdash s_i$
stands for $s_i$ can be parsed as $G_i$):
\begin{equation}\label{eq:reconf}
\begin{array}{ccccccccccc}
  G_0 & \to & G_1 & \to & \cdots & \to & G_{k-1} & \to & G_k & \to & \cdots
  \\
  \intercal &  & \intercal &  & \cdots & & \intercal & & \intercal & & \cdots
  \\
  s_0 & \stackrel{r_1}{\leadsto} & s_1 & \stackrel{r_2}{\leadsto} & \cdots & \stackrel{r_{k-1}}{\leadsto} & s_{k-1} & \stackrel{r_k}{\leadsto} & s_k & \stackrel{r_{k+1}}{\leadsto} & \cdots
\end{array}
\end{equation}
We assume that most of the run-time reconfigurations produce graphs
that do not violate $\inv$.
Occasionally, the graph obtained by a run-time reconfiguration, say
$G_i$, may violate $\inv$.
Our approach essentially computes how to rewrite graph $G_i$ to 
a graph $G_{i+1}$ satisfying $\inv$ and then reflect this into $s_i$ by means
of reconfigurations leading to a state $s_{i+1}$ with architecture $G_{i+1}$.

We propose a simple methodology that can select a production
that when applied to $G_i$ induces a reconfiguration of the violating
system into a state whose style satisfies $\inv$.
We assume a monitoring mechanism that triggers our methodology
whenever a reconfiguration yields to an invalid system.

Once, the productions and an architectural invariant $\inv$ yielding
the architectural style of interest are established (as done in
Example~\ref{ex:style}), our methodology consists of the following
steps:
\begin{enumerate}
\item \label{mth:parse} The architecture (say $G$) corresponding to
  the configuration of the current system is computed through ADR
  parsing.
\item \label{mth:check} Check that $G$ satisfies $\inv$.
\item \label{mth:pre} If $G \nvDash \inv$ then, for each production
  $p$, compute the weakest pre-condition $\phi$ wrt $\inv$.
\item\label{mth:last} Select a production $p$ (if any) such that $G
  \models \phi$ and apply it to $G$ to determine the reconfiguration
  needed for the system to reach a valid state.
\end{enumerate}

In step~\ref{mth:parse}, we rely on the \emph{parsing} mechanism of
ADR (cf.~\cite{bllmt08}) whereby productions can be used ``backward''
to retrieve the architecture of a configuration. For space limit, we do
not present the parsing mechanism and refer the interested reader
to~\cite{bllmt08}.
In step~\ref{mth:check}, we assume that an underlying monitoring
mechanism uses the $\models$ relation of our logic to determine if the
graph $G$ computed in step~\ref{mth:parse} violates the invariant.
In such case, step~\ref{mth:pre} uses the algorithm on each
production to compute their weakest preconditions (this step does
not need to be re-iterated at each reconfiguration).
Finally, in step~\ref{mth:last}, if the architecture of the violating
system satisfies one of the computed preconditions, such production is
a candidate to establish a new architecture and trigger the appropriate
reconfigurations on the invalid system.
Note that the morphism that invalidate $G \models \inv$ indicates
which part of the system has to be rewritten, while the production $p$
suggests plausible reconfigurations.

In\secref{sec:casestudy} we apply the methodology above to a small
example.


\section{Applying the methodology}
\label{sec:casestudy}
We consider a scenario where a flight search engine allows users to
book flights.

First, we use the type graph in Example~\ref{ex:tgandg} to model our
scenario in ADR.
Note that, in the type graph of Example~\ref{ex:tgandg}, there is only
one type of node $\chainport$ while the types of edges are $\client$
(for clients), $\bookF$ (for the booking flights services), $\findF$
(for the broker service finding flights), $\flights$ (for the
different flights available), $\flight$ (for the flight to be booked),
and $\pay$ and $\fpay$ (for completed or failed payment services,
respectively).
Consider the following productions:
\[
  \begin{array}{c@{\hspace{0cm}}c@{\hspace{0cm}}}
    \begin{prodenv}
      $\findflightprod$
      \\
      \xymatrix@C=.25cm@R=.25cm{
        &
        \bookF
        \\
        \chainport \ar@{.}[r] & \stackrel{u_2}{\chainport} &
        \taritytwo{\findFobj{ff}}{r}{l} &
        \stackrel{u}{\chainport} &
        \ntaritytwo{\flightsobj{fs}}{r}{l}&
        \stackrel{u_1}{\chainport} & \chainport \ar@{.}[l]
        \save "1,2"."2,6"*+[F.] \frm{} \restore
      }
    \end{prodenv}
    &
   \begin{prodenv}
      $\bookflightprod$
      \\
      \xymatrix@C=.2cm@R=.2cm{
        &
        \flights
        \\
        \chainport \ar@{.}[r] & \stackrel{u_2}{\chainport} &
        \ntaritytwo{\ \ \flightobj{f} \ \ }{r}{l} &
        \stackrel{u}{\chainport} &
        \ntaritytwo{\ \ \payobj{p}\ \ }{r}{l}&
        \stackrel{u_1}{\chainport} & \chainport \ar@{.}[l]
        \save "1,2"."2,6"*+[F.] \frm{}
        \restore }
    \end{prodenv}
    \\[3pc]
    \begin{prodenv}
      $\browseflightsprod$
      \\
      \xymatrix@C=.3cm@R=.3cm{
        &
        \flights
        &
        & \ntedge{\flightsobj{f_1}} \ar@(r,u)[dr] \ar@{-}@(l,u)[dl]
        \\
        \chainport \ar@{.}[rr] & & \stackrel{u_2}{\chainport} 
        & \ntaritytwo{\flightsobj{f_2}}{r}{l}
        & \stackrel{u_1}{\chainport}
        & \chainport \ar@{.}[l]
        \save "1,2"."2,5"*+[F.] \frm{} \restore
      }
    \end{prodenv}
    &
    \begin{prodenv}
      $\noFsprod$
      \\
      \xymatrix@C=.25cm@R=.25cm{
        &{\flights}\\
        \chainport \ar@{.}[r] & \stackrel{u_2}{\chainport} & \stackrel{u_1}{\chainport}
        & \chainport \ar@{.}[l]
        \save "1,2"."2,3"*+[F.] \frm{} \restore }
    \end{prodenv}
    \qquad
    \begin{prodenv}
      $\delFprod$
      \\
      \xymatrix@C=.25cm@R=.25cm{
        &{\flight}\\
        \chainport \ar@{.}[r] & \stackrel{u_2}{\chainport} & \stackrel{u_1}{\chainport}
        & \chainport \ar@{.}[l]
        \save "1,2"."2,3"*+[F.] \frm{} \restore }
    \end{prodenv}
  \end{array}
\]
where $\findflightprod$ establishes a broker service $\findF$,
$\bookflightprod$ yields a flight ($\flight$) connected to a payment
service ($\pay$), $\browseflightsprod$ generates as many flights as
necessary, and finally $\delFprod$ and $\noFsprod$ respectively remove
and stop adding flights to the design.

Services can either be composed with other services using
$\findflightprod$ and $\bookflightprod$ like for instance when one
chooses a specific flight and the system needs to ``invoke'' another
service (payment service) to complete the request,
or branch using the production $\browseflightsprod$ to represent
the different flights a customer can choose from.

\begin{figure}[t]\centering\tiny
  $\begin{array}{c @{\hspace{2cm}} c}
    \xymatrix@C=.3cm@R=.3cm{
      &&&& \ntedge{\flightobj{f_1}} \ar@{-}@(l,ur)[dl] \ar@(r,l)[r]
      & \stackrel{z}{\chainport}
      & \ntedge{\payobj{p}} \ar@{-}@(l,r)[l] \ar@(ur,u)[dlllll]
      \\
      \tarityone{\clientobj{c}}{r} &
      \stackrel{v}{\chainport} &
      \tedge{\findFobj{ff}} \ar@{-}@(l,r)[l] \ar@(r,l)[r] 
      &\stackrel{w}{\chainport}
      & \ntedge{\flightsobj{f_n}} \ar@{-}@(l,r)[l] \ar@(l,dr)[lll]
      &
    }
    &
    \xymatrix@C=.3cm@R=.3cm{
      &&&& \ntedge{\flightobj{f_1}} \ar@{-}@(l,ur)[dl] \ar@(r,l)[r]
      & \stackrel{z}{\chainport}
      & \ntedge{\fpayobj{pf}} \ar@{-}@(l,r)[l] \ar@(ur,u)[dlllll]
      \\
      \tarityone{\clientobj{c}}{r} &
      \stackrel{v}{\chainport} &
      \tedge{\findFobj{ff}} \ar@{-}@(l,r)[l] \ar@(r,l)[r] 
      &\stackrel{w}{\chainport}
      & \ntedge{\flightsobj{f_n}} \ar@{-}@(l,r)[l] \ar@(l,dr)[lll]
      &
    }
    \\[3pc]
    \text{(a)} & \text{(b)}
  \end{array}$
  \caption{\label{fig:scenario}A simple scenario}
\end{figure}
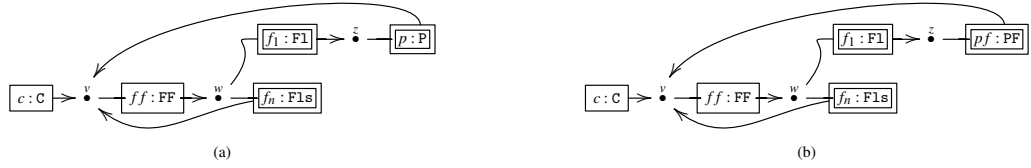

Figure~\ref{fig:scenario}(a) shows the architectural style of a system
where a client books a flight and successfully pays for it.
Initially, the client searches for a flight by invoking the
$findFlight$ service which, in turn, invokes different airlines about
their flights.
Once a flight is selected a payment service is used to complete the
transaction.

Sometimes, failures are possible during the payment; this is modelled
in Figure~\ref{fig:scenario}(b) where the payment edge $\pay$
reconfigures as an $\fpay$ edge.
We show how to apply our methodology in this scenario.

\newcommand{\invcs}{\phi_\text{\flight}}

The style we consider consists of the productions above and the invariant
\[ \invcs \  = \ \exists \flight ( \lvar{x_1}, \lvar{x_1'}) .\
\exists \pay ( \lvar {x_2'}, \lvar{x_2}) . \ \lvar{x_1} = \lvar{x_2} \]
that specifies that some flight $\flight$ has to be connected to a
successful payment $\pay$.

Following the methodology presented in\secref{sec:methodology}, we
need to check if graph $G_b$ given in Figure~\ref{fig:scenario}(b)
satisfies the invariant $\invcs$ and find that $G_b \nvDash \invcs$.
In fact, there is no edge of type $\pay$ in $G_b$
so we invoke $\wpre {\bar h} h p
\invcs$ on every production $p$ where $h$ is $\emptyset$ (since
$\invcs$ is a closed formula) and $\bar h$ maps the interface nodes of
$p$.
We have $\wpre {\bar h} \emptyset p \invcs = \invcs$ for all $p \neq
\bookflightprod$ whereas, for $p = \bookflightprod$, $\wpre {\bar h}
\emptyset p \invcs = \true$.

We show that $\wpre {\bar h} \emptyset p \invcs$ acts in the same way
(and yields $\invcs$) for any $p \neq \bookflightprod$ since such
productions do not have edges of type $\flight$ or $\pay$ in their RHS.
We have to compute $\wDef {p'} {\emptyset} \invcs \land \wp
{p',\bar h} {\emptyset, \envempty} \invcs$ by first applying the case
of existential quantification (cf. Definition~\ref{def:wp}):
\[ 
\big(\ \bigvee_{j=1, \dots , 5}
\wDef p {\envirWp_{j}} {\invcs'} \ \big)
\quad \land \quad
\big( \ \bigvee_{j=1, \dots , 5}
\exists \flight (\lvar{x_1}, \lvar{x_1'}).\wp {p,\bar h} {\emptyset, \envirWp_{j}} {\invcs'}
\lor \wDef p {\envirWp_{j}} {\invcs'}
\ \big)
\]
\noindent
where $\invcs' = \exists \pay ( \lvar {x_2'}, \lvar{x_2}) . \ \lvar{x_1} = \lvar{x_2}$.
Let $v_1$ and $v_2$ be representative nodes outside the
RHS of the productions above, $u_1$ and $u_2$ be interface nodes
of the productions.
The assignments
\begin{eqnarray*}
  \envirWp_{1  } & = & \{\ \lvar x_1 \mapsto (\exists , \flight, u_1),
  \ \ \lvar x_1' \mapsto (\exists , \flight, v_1) \ \}
  \\
  \envirWp_{2  } & = & \{\ \lvar x_1 \mapsto (\exists , \flight, u_2),
  \ \ \lvar x_1' \mapsto (\exists , \flight, v_1) \ \}
  \\
  \envirWp_{3  } & = & \{\ \lvar x_1 \mapsto (\exists , \flight, v_1),
  \ \ \lvar x_1' \mapsto (\exists , \flight, u_1) \ \}
  \\
  \envirWp_{4  } & = &  \{\ \lvar x_1 \mapsto (\exists , \flight, v_1),
  \ \ \lvar x_1' \mapsto (\exists , \flight, u_2) \ \}
  \\
  \envirWp_{5  } & = & \{\ \lvar x_1 \mapsto (\exists , \flight, v_1),
  \ \ \lvar x_1' \mapsto (\exists , \flight, v_2) \ \}
\end{eqnarray*}
are the only ones to consider for the first quantification.
Instead, for the other existential quantification $\exists \pay(\lvar
{x_2'} , \lvar {x_2})$ yields
\[
\big( \bigvee_{j,k=7,\dots , 11}
\wDef {p'} {\envirWp_{j} \cup \envirWp_{k}}{\invcs''}
\big)
\quad \land \quad
\big( \bigvee_{j,k=7, \dots , 11}
\exists \flight(\lvar {x_1}, \lvar {x_1'}).\exists \pay(\lvar {x_2'} , \lvar {x_2}).
\wp {p',\bar h} {\emptyset, \envirWp_{j} \cup \envirWp_{k}}{\invcs''}
\lor \wDef {p'} {\envirWp_{j} \cup \envirWp_{k}}{\invcs''}
\big)
\]
where $\invcs''$ is the equality $\lvar{x_1} = \lvar{x_2}$ and the
assignments $\envirWp_7,\ldots,\envirWp_{11}$ are:
\begin{eqnarray*}
  \envirWp_{7  } & = & \{\ \lvar x_2 \mapsto (\exists , \pay, u_1),
  \ \ \lvar x_2' \mapsto (\exists , \pay, v_1) \ \}
  \\
  \envirWp_{8  } & = & \{\ \lvar x_2 \mapsto (\exists , \pay, u_2),
  \ \ \lvar x_2' \mapsto (\exists , \pay, v_1) \ \}
  \\
  \envirWp_{9  } & = & \{\ \lvar x_2 \mapsto (\exists , \pay, v_1),
  \ \ \lvar x_2' \mapsto (\exists , \pay, u_1) \ \}
  \\
  \envirWp_{10  } & = &  \{\ \lvar x_2 \mapsto (\exists , \pay, v_1),
  \ \ \lvar x_2' \mapsto (\exists , \pay, u_2) \ \}
  \\
  \envirWp_{11 } & = & \{\ \lvar x_2 \mapsto (\exists , \pay, v_1),
  \ \ \lvar x_2' \mapsto (\exists , \pay, v_2) \ \}
\end{eqnarray*}
Finally, applying the case for node equality in the auxiliary map 
$\fwp p {\psi_1, \psi_2,\psi_3} {\lvar x_1 = \lvar x_2}{\envirWp}$ of
Definition~\ref{def:wp}, we get
\begin{eqnarray}\label{eq:wd1}
  \bigvee_{j,k}
  \wDef p {\envirWp_{j} \cup \envirWp_{k}}{\invcs''}
  & \quad = \quad & 
  \true \lor \true \lor \dots = \true
  \\
  \label{eq:wp1}
  \bigvee_{j,k}
   \exists \flight(\lvar{x_1}, \lvar{x_1'}). \exists \pay(\lvar{x_2'} , \lvar{x_2}).
   \wp {p,\bar h} {\emptyset, \envirWp_{j} \cup \envirWp_{k}}{\invcs''}
  & \quad = \quad &
  ( \invcs  \; \lor \; \false) \; \lor \; ( \invcs \; \lor \; \false )
   \; \lor \; \dots \; = \; \invcs
\end{eqnarray}
which yield $\wpre {\bar h} \emptyset p \invcs$ since~\eqref{eq:wd1}
and~\eqref{eq:wp1} respectively correspond to $\wDef p {\emptyset}
\invcs$ and $\wp {p,\bar h} {\emptyset, \envempty} \invcs$.

We now consider $p = \bookflightprod$ and show that $\wpre {\bar h}
\emptyset p \invcs = \true$.
As in the previous case, we consider the quantifications for which we
have to consider the extra mappings due to $\flight$ and $\pay$:
\begin{eqnarray*}
  \envirWp_6 & = & \curlybraces{ \lvar {x_1} \mapsto (\exists ,
  \flight, u), \lvar {x_1'} \mapsto (\exists , \flight, u_2) }
\\
\envirWp_{12} & = & \curlybraces{ \lvar {x_2'} \mapsto (\exists , \pay,
  u_1), \lvar {x_2} \mapsto (\exists , \pay, u) }
\end{eqnarray*}
where $u_1$ and $u_2$ are the production's interface nodes as
before and $u$ is its unique internal node.
By the quantification cases we have
\[
\big( \bigvee_{j,k}
\wDef p {\envirWp_{j} \cup \envirWp_{k}}{\invcs''}
\big)
\quad \land \quad
\big( \bigvee_{j,k}
\exists \flight(\lvar{x_1}, \lvar {x_1'}).\exists \pay(\lvar {x_2'}, \lvar{x_2}).
\wp {p,\bar h} {\emptyset, \envirWp_{j} \cup \envirWp_{k}}{\invcs''}
\lor \wDef p {\envirWp_{j} \cup \envirWp_{k}}{\invcs''}
\big)
\]
where $j=1,\dots , 6$ and $k=7, \dots , 12$.

Finally, applying the case for node equality in the auxiliary map 
$\fwp p {\psi_1, \psi_2,\psi_3} {\lvar x_1 = \lvar x_2}{\envirWp}$ of
Definition~\ref{def:wp}, we get
\begin{eqnarray}\label{eq:wd2}
  \bigvee_{j,k}
  \wDef p {\envirWp_{j} \cup \envirWp_{k}}{\invcs''}
  & \quad = \quad & 
  \true \lor \true \lor \dots = \true
  \\
  \label{eq:wp2}
  \bigvee_{j,k}
   \exists \flight(\lvar{x_1}, \lvar{x_1'}). \exists \pay(\lvar{x_2'} ,
   \lvar{x_2} )\wp {p,\bar h} {\emptyset, \envirWp_{j} \cup \envirWp_{k}}{\invcs''}
   & \quad = \quad &
   ( \exists \flight(\lvar{x_1}, \lvar{x_1'}). \exists \pay(\lvar{x_2'} ,
   \lvar{x_2} ).\true \: \lor \: \true ) \: \lor \: \dots \: = \: \true
\end{eqnarray}
Note that the weakest pre-conditions is the conjunction
of~\eqref{eq:wd2} and~\eqref{eq:wp2}, that is $(\wDef p {\envempty}
\invcs \land \wp {p,\bar h} {\emptyset, \envempty} \invcs) = \true$

The next step requires that we check whether the graph $G_b$ given in
Figure~\ref{fig:scenario}(b) satisfies any of the weakest
pre-conditions computed.  $G_b \nvDash \exists \flight ( \lvar {x_1},
\lvar {x_1'}) .  \exists \pay ( \lvar {x_2'}, \lvar {x_2}) . \lvar {x_1}
= \lvar {x_2}$ but instead $G_b \models \true$ and therefore we know that
by applying the production $\bookflightprod$ we get a graph $G_b'$ that
satisfies the invariant $\invcs$.


\section{Related work}
\label{sec:rw}
Formal approaches based on architectural styles to control
architectural reconfigurations have been proposed, among other,
in~\cite{HirschIM99,Allen98FASE,Metayer98,bllmt08}.
In those proposals reconfigurations are typically applied uniformly
across the design.
For instance, in~\cite{Metayer98,bllmt08} graph grammars and hyper-edge
replacements are used to represent styles in terms of graph
configurations freely generated by some productions (and it is not
easy to specify conditions to extract subsets of such
graph-languages).

Our work mitigates this effect by means of asserted productions that
provide a finer control on the applicability conditions as done in
other graph-transformation approaches.
For instance, our approach is similar to the one in~\cite{HabelPR06}
where graph programs are extended to programs over high-level rules
with application conditions; on such programs weakest pre-conditions
can be defined automatically.
Nevertheless,~\cite{HabelPR06} aims at verifying computational
properties of systems rather than architectural ones and does
that in a different way only after generating the various
state systems.
In~\cite{DBLP:journals/computer/GarlanCHSS04} constraints on the architecture
are used to guarantee invariants of systems.
More precisely, reconfigurations can occur only if such constraints are
not violated.
This is not always realistic in open systems, therefore they do not impose
limitations on run-time reconfigurations and search for new
reconfigurations that can lead the system in a desired state.

In~\cite{DBLP:conf/tacas/CobleighGP03} an assume-guarantee mechanism
is adopted to provide a learning algorithm which provides an assumption
satisfying a sufficient condition in order for the component to guarantee
the given invariant. This is achieved by model checking every component
of the system against an invariant.
This is similar to the weakest pre-condition we present in this paper 
but instead of computing the weakest assumption for every component 
of the system we compute the weakest pre-condition for every
design production.
We can later use our algorithm for applying the methodology described
in \S~\ref{sec:methodology} for identifying the possible design production(s)
(if any) to aid in fixing the architectural violation of the system.

In~\cite{DBLP:conf/icse/BeckerBGKS06} the authors present an approach
for designing safe systems by inspecting whether certain reconfigurations
can lead to invalid graphs that represent invalid systems.
This is achieved by verifying that the backward application of
reconfigurations to a forbidden graph pattern cannot lead to a graph
pattern representing a safe system (a set of forbidden graph patterns
model an invariant).
This method can provide a safe system in the sense that it cannot
lead to a state that violates a structural invariant by the use of
reconfigurations but it is very complex to handle unexpected system
failures.

In~\cite{DBLP:conf/fase/EhrigERBP10} self-healing systems
are modelled by specifying different types of rules; for the ideal
system behaviour, for different predictable failures and for fixing
the different failures identified earlier. This approach is different
to what we propose in this paper as they design the rules according to
the misbehaviours they expect at run time and do not necessarily
handle unexpected failures or changes of the system.

Different approaches to specify self-managing systems are surveyed in
\cite{DBLP:conf/woss/BradburyCDW04}.
The authors group the different approaches according to their ability
to select different reconfigurations that should occur to re-establish
a correct state.
They present three type of selections namely, called \emph{pre-defined
  selection} (a reconfiguration is chosen prior to the execution based
on a pre-defined selection), \emph{constrained selection from a
  pre-defined set} (a reconfiguration designed for the given situation
is chosen) and \emph{unconstrained selection} (unconstrained choice
regarding the appropriate change to make).
All the approaches presented in the survey lie in either of the former
two categories and according to~\cite{DBLP:conf/woss/BradburyCDW04},
none of the approaches survyed falls in the unconstrained selection
category.
Our approach does not lie neither in the pre-defined nor in
constrained selection categories.
It is not clear to us if our approach can be considered an
unconstrained selection.
In fact, we do not choose the reconfigurations to apply according to
the misbehaviours expected at run time.
Instead we use our weakest pre-condition algorithm to identify which
of the existing configurations (not designed for the specific
violation) can re-establish the architectural style of our system.
We remark that most of the rules given at design time typically are
meant to specify the architectural style of a system, not its
misbehaviours (for instance, in ADR this might be addressed with
reconfiguration rules rather than productions).
However, even if some productions were introduced to tackle (or
prevent) some misbehaviours, our approach enables such rules to be
used also for unexpected violations.


\section{Conclusion and future work}
\label{sec:conc}
We introduced a methodology inspired by Design by Contract 
(DbC)~\cite{DbC92} to guarantee properties of architectural designs.
Technically this is achieved by ($i$) equipping ADR with a
logic tailored to express such properties and ($ii$) devising
an algorithm to compute weakest pre-conditions for ADR productions.

Albeit very simple, our logic can express rather interesting properties
(cf. Example~\ref{ex:interesting}).
It allows us to improve the expressiveness of ADR and to specify interesting
properties exploiting the 'hierarchical nature' of ADR graphs. This paper is
a first step in the exploration of the use of DbC in
architectural style reconfigurations.

Using our methodology we can fix architecturally our graphs, provided
that we have the appropriate productions to do this.
Currently, our methodology works if there is a single production for recovering
a failure, but we see this work as a first step towards the more realistic situation
where to tackle failures one tries to apply a number of productions.
More precisely, one could compute a sequence of productions by iterating the
methodology in \S~\ref{sec:methodology} on the weakest pre-condition obtained
at every ``round'' (starting from the invariant) until either ‘false’ or a
valid style is reached. We note that this opens other interesting questions.
For example, when different sequences of productions are found, one could devise
criteria to order them, or else to try to find criteria for good or best strategies. 
Generalising our idea for computing 'strategies' based on many productions to
recover failures could be a a very interesting future direction.

We expect such research to lead to extensions of the logic and also
like stated earlier extensions to the methodology to be able to handle
more complex violations that might require more design productions to
fix a system's architecture.


\paragraph{Acknowledgements} The authors thank Andrea Vandin for
valuable comments and suggestions.

\bibliographystyle{eptcs}
\bibliography{mybibliography}

\end{document}